\documentclass[aps,showpacs,amsmath,amssymb,superscriptaddress,prl,preprint]{revtex4-1}

\usepackage{graphicx}
\usepackage{bm}
\usepackage{color}
\usepackage{epstopdf}
\usepackage[version=3]{mhchem}

\begin{document}

\title{{Current-induced magnetization switching in a chemically disordered A1 CoPt single layer}}

\author{Zehan Chen} 
\affiliation{College of New Materials and New Energies, Shenzhen Technology University, Shenzhen 518118, China}


\author{Lin Liu} 
\affiliation{College of New Materials and New Energies, Shenzhen Technology University, Shenzhen 518118, China}

\author{Zhixiang Ye} 
\affiliation{College of New Materials and New Energies, Shenzhen Technology University, Shenzhen 518118, China}

\author{Zhiren Chen} 
\affiliation{College of New Materials and New Energies, Shenzhen Technology University, Shenzhen 518118, China}

\author{Hongnan Zheng} 
\affiliation{College of New Materials and New Energies, Shenzhen Technology University, Shenzhen 518118, China}


\author{Wei Jia} 
\affiliation{College of New Materials and New Energies, Shenzhen Technology University, Shenzhen 518118, China}

\author{Qi Zeng} 
\affiliation{College of New Materials and New Energies, Shenzhen Technology University, Shenzhen 518118, China}

\author{Ning Wang} 
\email{wangning@sztu.edu.cn}
\affiliation{College of New Materials and New Energies, Shenzhen Technology University, Shenzhen 518118, China}

\author{Boyuan Xiang} 
\affiliation{College of New Materials and New Energies, Shenzhen Technology University, Shenzhen 518118, China}

\author{Tao Lin} 
\affiliation{College of New Materials and New Energies, Shenzhen Technology University, Shenzhen 518118, China}

\author{Jing Liu} 
\affiliation{College of New Materials and New Energies, Shenzhen Technology University, Shenzhen 518118, China}

\author{Mingxia Qiu} 
\affiliation{College of New Materials and New Energies, Shenzhen Technology University, Shenzhen 518118, China}


\author{Shunpu Li} 
\affiliation{College of New Materials and New Energies, Shenzhen Technology University, Shenzhen 518118, China}

\author{Ji Shi} 
\affiliation{School of Materials and Chemical Technology, Tokyo Institute of Technology, Tokyo 152-8552, Japan}

\author{Peigang Han}
\email{hanpeigang@sztu.edu.cn} 
\affiliation{College of New Materials and New Energies, Shenzhen Technology University, Shenzhen 518118, China}

\author{Hongyu An}
\email{anhongyu@sztu.edu.cn}
\affiliation{College of New Materials and New Energies, Shenzhen Technology University, Shenzhen 518118, China}

\begin{abstract}

We report the first demonstration of the current-induced magnetization switching in a perpendicularly magnetized A1 CoPt single layer. We show that good perpendicular magnetic anisotropy can be obtained in a wide composition range of the A1 Co$_\text{1-x}$Pt$_\text{x}$ single layers, which allows to fabricate perpendicularly magnetized CoPt single layer with composition gradient to break the inversion symmetry of the structure. By fabricating the gradient CoPt single layer, we have evaluated the SOT efficiency and successfully realized the SOT-induced magnetization switching. Our study provides an approach to realize the current-induced magnetization in the ferromagnetic single layers without attaching SOT source materials.

\end{abstract}

\maketitle


Current-induced magnetization switching through spin-orbit torques (SOTs) is essential for the spin-orbitronic-based memory devices with low-energy consumption and high speed~\cite{miron2011perpendicular,miron2010current,liu2012spin,liu2012current,yu2014switching,fan2014magnetization,cubukcu2014spin,woo2014enhanced,zhao2015spin,garello2014ultrafast,liu2014control,yu2014magnetization,wang2014spin}. Typically, in a heavy metal (HM)/ferromagnet (FM) bilayer, the SOTs can be generated through the bulk spin Hall effect in the HM layer and the Rashba effect at the HM/FM interface~\cite{akyol2015effect,yu2014current,haogiant,garello2013symmetry,avci2012magnetization,kim2014anomalous,PhysRevB.91.214434,zhang2015spin,avci2014fieldlike,hayashi2014quantitative}. By applying a charge current, the generated SOTs transfer into the FM layer, and act as effective magnetic fields to manipulate the magnetization of the FM layer. Therefore, it was considered that the HM or other SOT source materials were indispensable for generating SOTs and manipulating the magnetization of the FM layer. However, recent several studies show that the current-induced magnetization switching can be realized in a FM  single layer without attaching the HM layer~\cite{tang2020bulk,liu2020electrical,zhang2020current,lee2020spin}. For instance, Tang $et$ $al$. reported the current-induced magnetization switching in an ordered L1$_\text{0}$ FePt single layer~\cite{tang2020bulk}. They found that the inherent structural gradient along the film normal direction breaks the structure symmetry, which can generate sizable SOTs and responsible for the magnetization switching. Liu $et$ $al$. also reported the current-induced magnetization switching in the L1$_\text{0}$ FePt single layer due to the composition gradient~\cite{liu2020electrical}. Zhang $et$ $al$. reported the current-induced magnetization switching in a CoTb amorphous single layer~\cite{zhang2020current}. They interpreted that the local inversion broken symmetry inside the layer can generate net SOTs and responsible for the magnetization switching. Lee $et$ $al$. studied the SOT generations in the CoTb single layers~\cite{lee2020spin}. They interpreted that the bulk spin-orbit interaction within the CoTb layer plays a major role for the SOT generation. Very recently, Zhu $et$ $al$. reported the SOT generation in a chemically disordered A1 CoPt single layer with in-plane magnetic anisotropy~\cite{zhu2020observation}. They interpreted that the SOTs are likely generated by the spin Hall effect in the CoPt layer, since no long-range asymmetry was observed. However, in their study, nonzero dampinglike torque only exists above 8 nm and no current-induced magnetization switching was studied. So far, the understanding of the SOT generation and current-induced magnetization switching in the FM single layers still remains elusive and is just beginning to be probed.

In this work, we for the first time report the current-induced magnetization switching in a perpendicularly magnetized A1 CoPt single layer. We show that good perpendicular magnetic anisotropy can be obtained in a wide composition range of the A1 Co$_\text{1-x}$Pt$_\text{x}$ single layers, which allows to fabricate perpendicularly magnetized CoPt single layer with composition gradient to break the inversion symmetry of the structure. By fabricating the gradient CoPt single layer, we have evaluated the SOT generation efficiency and successfully realized the SOT-induced magnetization switching.


For the sample fabrication, Co$_\text{1-x}$Pt$_\text{x}$ single layer films were deposited on MgO (111) single crystal substrates at 350 $^\text{o}$C by magnetron sputtering. Co and Pt targets were co-sputtered with different powers to form Co$_\text{1-x}$Pt$_\text{x}$ with different composition ratios. The Co sputtering power was fixed as 30 W, and the Pt sputtering power was varied from 30 to 70 W. The concentration of Pt was calculated from the deposition rate at each sputtering power. The base pressure in the chamber before deposition was better than 1 $\times$10$^{-6}$ Pa, and the deposition pressure was 0.4 Pa. During the sputtering, argon gas was supplied. The film thickness for each composition ratio was controlled by the deposition time with a precalibrated deposition rate. Crystal structure of the films was characterized by x-ray diffraction (XRD) with Cu K$_{\alpha}$ irradiation, and a vibrating sample magnetometer (VSM) was used to measure the magnetic properties. For the electrical transport measurements, the substrates were patterned into a Hall bar shape with width of 20 $\mu$m and length of 100 $\mu$m.

Figures 1(a)-(e) demonstrate the anomalous Hall effect (AHE) hysteresis loops of 5-nm-thick Co$_\text{1-x}$Pt$_\text{x}$ single layers with different compositions. Well-defined, square AHE hysteresis loops were obtained, indicating good perpendicular magnetic anisotropy (PMA) of the Co$_\text{1-x}$Pt$_\text{x}$. As can be seen, by increasing the Pt concentration, the AHE resistance of the Co$_\text{1-x}$Pt$_\text{x}$ decreases gradually, which confirms that the AHE signal is induced by the Co magnetization. It is noteworthy that the PMA is realized in a wide composition range from Co$_\text{0.62}$Pt$_\text{0.38}$ to Co$_\text{0.41}$Pt$_\text{0.59}$, which shows that CoPt alloy has good tunability to manipulate its compositions without sacrificing its PMA. Therefore, we can fabricate a CoPt single layer with composition gradient to break the inversion symmetry of the structure. A 5-nm-thick Co$_\text{1-x}$Pt$_\text{x}$ (x = 0.38$\to$0.59) single layer was fabricated. During the deposition, the Co sputtering power was fixed as 30 W, and the Pt sputtering power was continuously changed from 30 to 70 W by the sputtering controller. In the sputtering controller, by inputting the initial power 30 W, ending power 70 W and deposition time period, the power will vary linearly from 30 to 70 W. Hereafter, we define the Co$_\text{1-x}$Pt$_\text{x}$ (x = 0.38$\to$0.59) as gradient CoPt. As shown in Fig. 1(f), the 5-nm thick gradient CoPt single layer exhibits good PMA.

CoPt alloy has three main structural phases, which are A1 phase, L1$_1$ phase and L1$_0$ phase. The A1 phase is a face-centered cubic (fcc) structure with each position randomly occupied by Co or Pt atom~\cite{pan2019large}. The L1$_1$ phase is a rhombohedral structure with each atom layers alternately occupied by Co and Pt atoms along the [111] direction~\cite{gao2019formation}. The L1$_0$ phase is a face-centered tetragonal (fct) structure with each atom layers alternately occupied by Co and Pt atoms along the [100] direction~\cite{an2015perpendicular}. First, we can rule out the L1$_0$ phase in our Co$_\text{1-x}$Pt$_\text{x}$, since the formation of the L1$_0$ phase requests much higher deposition/annealing temperature ($\sim$700 $^\text{o}$C)~\cite{an2015perpendicular,tang2020bulk}. Moreover, only (001)-oriented L1$_0$ CoPt exhibits PMA. While, since we used MgO (111) substrates, the Co$_\text{1-x}$Pt$_\text{x}$ can only epitaxially grow in [111] direction. To investigate whether the crystal structure of the Co$_\text{1-x}$Pt$_\text{x}$ in our study is chemically disordered A1 phase or ordered L1$_1$ phase, we conducted the XRD measurement.   As shown in Fig. 2, besides the MgO (111) peak, a peak located at around 41$^\text{o}$ can be observed. This peak can be indexed as the A1 CoPt (111) peak or L1$_1$ CoPt (222) peak~\cite{gao2019formation}. However,  the representative L1$_1$ CoPt (111) peak, which is located at around 21$^\text{o}$~\cite{gao2019formation}, can not be observed. Thus, we can confirm that the Co$_\text{1-x}$Pt$_\text{x}$ is A1 phase, and the peak at around 41$^\text{o}$ is A1 CoPt (111) peak. It is noteworthy that by increasing the Pt concentration, the A1 CoPt (111) peak gradually shifts to the smaller angle, which is well consistent with its composition, since the Pt (111) standard peak is located at around 40 $^\text{o}$~\cite{gao2019formation}. The PMA in our A1 Co$_\text{1-x}$Pt$_\text{x}$ single layer is induced by the magnetoelastic anisotropy. Since the lattice constants of the MgO (111) substrate and CoPt (111) crystal are 2.978 $\text{\AA}$ and 2.692 $\text{\AA}$, internal tensile stress is induced in the CoPt layer due to the large lattice mismatch (9.6$\%$), which favors the PMA~\cite{pan2019large}.

In the following, we evaluate the SOT generations in the uniform CoPt and gradient CoPt single layers. By applying an ac current, the second-harmonic AHE resistance was measured as a function of an in-plane external magnetic field using lock-in amplifiers, as shown in Fig. 3. The measured data was fitted by the formula $R_{2\omega}$ = -$R_\text{H}$$H_\text{L}$/2($H_\text{x}$ - $H_\text{K}$) in the data range that the magnetization is totally aligned along the external magnetic field~\cite{fan2014magnetization}. $R_\text{H}$ is the out-of-plane saturation AHE resistance, $H_\text{L}$ is the effective SOT field induced by the ac current, and $H_\text{K}$ is the magnetic anisotropy field. $H_\text{L}$ is obtained as 0.73  $\times$ 10$^{-11}$ Oe A$^{-1}$ m$^2$ in the Co$_\text{0.50}$Pt$_\text{0.50}$ and 2.31 $\times$ 10$^{-11}$ Oe A$^{-1}$ m$^2$ in the gradient CoPt single layers. Furthermore, the SOT generation efficiency can be estimated by ${\xi _{{\rm{}}}} =  {\mu_0{M_{\rm{s}}}} {t}H_\text{L}{2e/\hbar }J_\text{c} $, where $M_{\rm{s}}$, $t$, $\hbar$ and $J_\text{c} $ are the saturation magnetization, CoPt thickness, Planck's constant and current density, respectively. ${\xi _{\rm{}}}$ is calculated to be 0.009 for the Co$_\text{0.50}$Pt$_\text{0.50}$ and 0.028 for the gradient CoPt. The SOT efficiency in gradient CoPt is more than three times large of that in the Co$_\text{0.50}$Pt$_\text{0.50}$, indicating that the composition gradient breaks the inversion symmetry of the structure and results in a much larger SOT generation. This is consistent with previous studies on L1$_0$ FePt~\cite{tang2020bulk,liu2020electrical} and amorphous CoTb~\cite{zhang2020current}.

Because of the existence of the sizable SOT in the gradient A1 CoPt, current-induced magnetization switching is expected in the A1 CoPt single layer. Thus, we measured $R_\text{H}$ by sweeping the in-plane dc current $I_\text{dc}$. For the measurement, the in-plane magnetic field $H_\text{x}$ was applied to break the rotational symmetry of the SOT (see Fig. 4(a)). As shown in Fig. 4(b), by applying nonzero $H_\text{x}$, the current switches the magnetization of the gradient CoPt single layer between up and down directions, and the polarity of the magnetization switching is reversed by reversing the direction of $H_\text{x}$, which is consistent with the magnetization switching induced by the SOT. We also conducted same measurement for the Co$_\text{0.50}$Pt$_\text{0.50}$ single layer. However, no magnetization switching can be observed (see Fig. 4(c)). This is well consistent with the SOT measurement that the small SOT efficiency is not sufficient to switch the magnetization in the CoPt single layers lack of inversion asymmetry.


In summary, we have demonstrated the current-induced magnetization switching in the perpendicularly magnetized A1 CoPt single layer. We show that good PMA can be obtained in a wide composition range which allows to fabricate perpendicularly magnetized CoPt single layer with composition gradient to break the inversion symmetry of the structure. It is found that the SOT efficiency in the gradient A1 CoPt is much larger than that in the uniform Co$_\text{0.50}$Pt$_\text{0.50}$ single layer. Thanks to the sizable SOT generation, the SOT-induced magnetization switching is successfully realized in the gradient A1 CoPt single layer. Our study provides an approach to realize the current-induced magnetization in the FM single layers without attaching SOT source materials.

\section{Acknowledgments}
\begin{acknowledgments}
Ze.C., L.L. and Z.Y. contributed equally to this work. 

This work was supported by the Guangdong Basic and Applied Basic Research Foundation (Grant No. 2019A1515110230), the National Natural Science Foundation of China (Grant No. 52001215), the Featured Innovation Project of the Educational Commission of Guangdong Province of China (Grant No. 2019KTSCX203), Natural Science Foundation of Top Talent of SZTU (Grant No. 2019208 and 2019106101006).
\end{acknowledgments}
\clearpage


%

\begin{figure}[tb]
\center\includegraphics[scale=1]{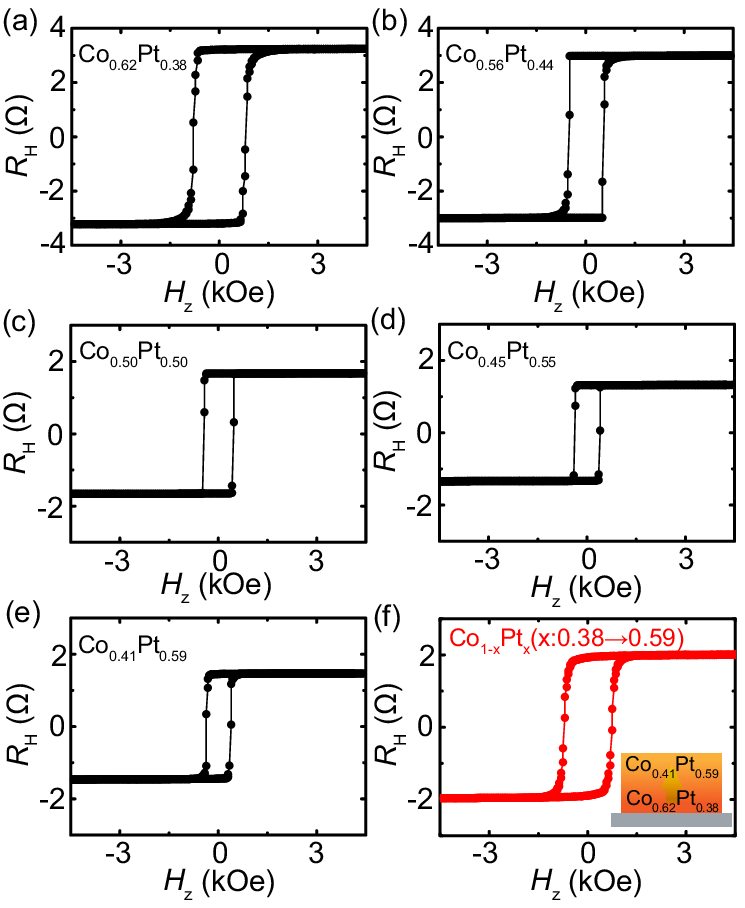}
\caption{The AHE resistance $R_\text{H}$ of the (a) Co$_{0.62}$Pt$_{0.38}$, (b) Co$_{0.56}$Pt$_{0.44}$, (c) Co$_{0.50}$Pt$_{0.50}$, (d) Co$_{0.45}$Pt$_{0.55}$, (e) Co$_{0.41}$Pt$_{0.59}$ and (f) gradient Co$_{1-x}$Pt$_{x}$ (x: 0.38$\to$0.59) single layers. The thickness of all the films are 5 nm. 
}
\label{fig1}
\end{figure}
\clearpage

\begin{figure}[tb]
\center\includegraphics[scale=1]{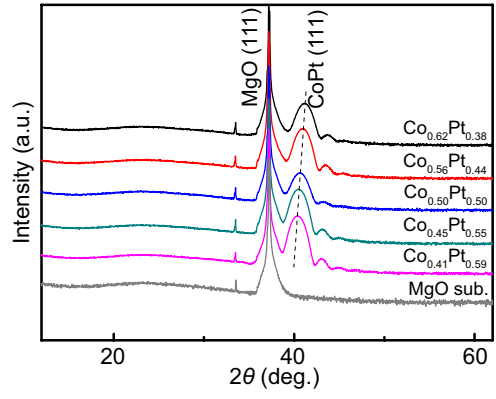}
\caption{XRD profiles of the CoPt single layer films with different compositions. The thickness of all the films are 5 nm.     
}
\label{fig2}
\end{figure}
\clearpage

\begin{figure}[tb]
\center\includegraphics[scale=1]{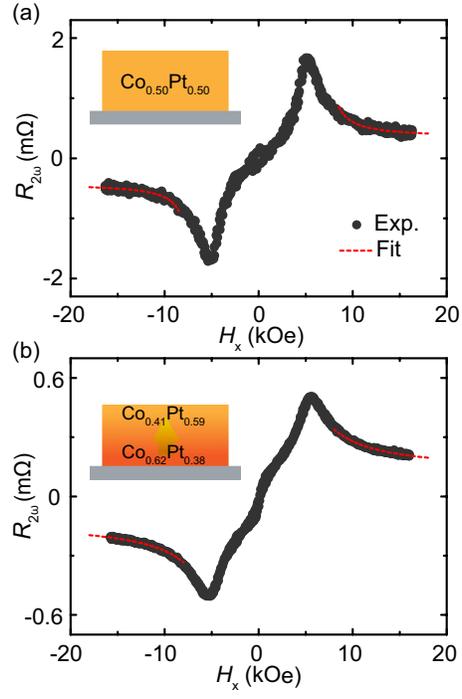}
\caption{
Second-harmonic AHE resistance as a function of the in-plane magnetic field for the (a) Co$_{0.50}$Pt$_{0.50}$ and (b) gradient CoPt single layer films.     
}
\label{fig3}
\end{figure}
\clearpage

\begin{figure}[tb]
\center\includegraphics[scale=1]{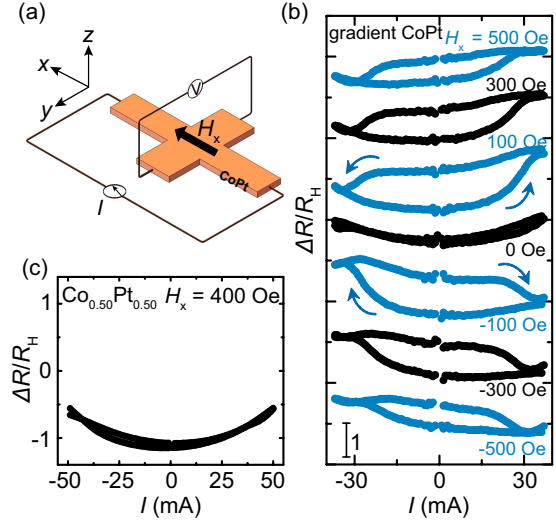}
\caption{
(a) Schematic of the setup for the current-induced magnetization switching measurement. Current-induced magnetization switching curves for the (b) gradient CoPt, (c) Co$_{0.50}$Pt$_{0.50}$ single layer films.  
}
\label{fig3}
\end{figure}
\clearpage

\end{document}